\documentclass{article}
\usepackage{graphicx}
\usepackage[hdivide={20mm,,20mm}, vdivide={15mm,,15mm}, nohead]{geometry}
\usepackage{setspace}
\usepackage{slashbox}
\usepackage{color}
\usepackage{multicol}
\usepackage{amsmath,amsthm,amssymb}
\usepackage[hang,flushmargin]{footmisc}
\usepackage{caption}
\usepackage{comment}
\usepackage{bm}
\makeatletter

\newenvironment{figurehere}
  {\def\@captype{figure}}
  {}
\makeatother

\makeatletter
\renewenvironment{thebibliography}[1]
{\section*{\refname\@mkboth{\refname}{\refname}}%
  \list{\@biblabel{\@arabic\c@enumiv}}%
       {\settowidth\labelwidth{\@biblabel{#1}}%
        \leftmargin\labelwidth
        \advance\leftmargin\labelsep
 \setlength\itemsep{-0.5pt}%
 \setlength\baselineskip{6pt}
        \@openbib@code
        \usecounter{enumiv}%
        \let\p@enumiv\@empty
        \renewcommand\theenumiv{\@arabic\c@enumiv}}%
  \sloppy
  \clubpenalty4000
  \@clubpenalty\clubpenalty
  \widowpenalty4000%
  \sfcode`\.\@m}
 {\def\@noitemerr
   {\@latex@warning{Empty `thebibliography' environment}}%
  \endlist}
\makeatother
 
\renewcommand{\refname}{{\normalsize References}}

\begin{document}

\begin{center}
\textbf{{\large  Random family method: Confirming inter-generational relations by restricted re-sampling}}
\vspace{3mm}

Takuma Usuzaki$^1$, Minoru Shimoyama$^1$, Shuji Chiba$^2$, Shuma Hotta$^3$

\vspace{3mm}
$^1Tohoku\ University\ School\ of\ Medicine$, $^2Dokkyo \ Medical \ University$, $^3Shibaura \ Institute \ of \ Technology$
\end{center}

\begin{multicols}{2}
Randomness is one of the important key concepts of statistics. Stuart A, Ord K and Arnold S\cite{ref1} mentioned that statistical hypothesis is a hypothesis  that is testable on the basis of observing a process that is modeled via a set of random variables. In epidemiology or medical science, we investigate our hypotheses and interpret results through this statistical randomness. Imposing some conditions to this randomness, interpretation of our result may be changed. In this article, we introduce the restricted re-sampling method to confirm inter-generational relations.
We call this method "Random family method".
\vspace{5mm}

\textbf{Definition}:
Let $X$ be a set. $x_n\in X$ for $n=1,2,\dots,$ then $(x_1,x_2,\dots)$ is called a sequence in $X$ and is denoted $\{x_i\}$.

\vspace{5mm}

\textbf{Definition}:
Let $f$ be a mapping. Let consider integer sequence $(1,2,\dots,n)$.$f:\mathbb{N}^n\rightarrow \mathbb{N}^n$,$f((1,2,\dots,n))=\{\{I_j\}|I_j \neq j,I_j,j\in\mathbb{N},1\leq I_j,j\leq n, I_j\neq I_k$ if $j\neq k\}$.

\vspace{5mm}

\textbf{Lemma}:
For $n>3$,  let $\{I_j\}$ be a image of $(1,2,\dots,n)$ by $f$. Let $N(n)$ be the number of $\{I_j\}$. $N(n)$ satisfies relation
\begin{eqnarray*}
N(n)=(n-1)(N(n-1)+N(n-2)).
\end{eqnarray*}

$\bm{Proof}$: $N(1)=0, N(2)=1, N(3)=2$. Let consider $N(n)$ for given $N(n-1)$ and $N(n-2)$. First, choosing a position of mapping $i(1\leq i\leq n)$, the number of selection is $n-1$. When $i$ is mapped $I_j$, let consider position of mapping $j$. If $j$ is mapped on $I_i$ the number of the residual mapping is $N(n-2)$, else if $j$ is mapped on another the number of residual mapping is $N(n-1)$.\qedhere

\vspace{5mm}

Let $\{I^N_j\}$ be  $N$-th $(1\leq N\leq N(n))$ image of $(1,2,\dots,n)$ by $f$. $\{x_{I^N_j}\}$ denotes $(x_{I^N_1}, x_{I^N_2},\dots, x_{I^N_n})$. Let $\{x_i\}$ and $\{y_i\}$ be sequence of $X$ and $Y$ respectively $(i=1,2,\dots,n)$. Besides a relation between $\{x_i\}$ and $\{y_i\}$, we can perform additional analyses by considering $\{x_{I^N_j}\}$  and $\{y_i\}$. In the following part of this article we present an example of this additional analysis. This is an application of permutation test or bootstrap method\cite{ref2}.

\vspace{5mm}
\textbf{Example}:\textbf{Coefficients}
\vspace{2mm}

Let $X$ and $Y$ have a bivariate normal distribution.

\vspace{2mm}
\textbf{Definition}:
Denote the means of $X$ and Y respectively by $\mu_1$ and $\mu_2$ and their respective variances by $\sigma_1^2$ and $\sigma_2^2$. The covariance of $(X,Y)$ is denoted by $cov(X,Y)$ and is defined by the expectation $cov(X,Y)=E[(X-\mu_1)(Y-\mu_2)]$.

\vspace{5mm}
\textbf{Definition}:
If each of $\sigma_1$ and $\sigma_2$ is positive, then the correlation coefficient between $X$ and $Y$ is defined by
\begin{eqnarray*}
\rho =\frac{E[(X-\mu_1)(Y-\mu_2)]}{\sigma_1\sigma_2}.
\end{eqnarray*}

\textbf{Definition}:
Given $X_1=x_1, X_2=x_2, \dots, X_n=x_n(n>2)$, mean least estimator of simple linear regression coefficient which denoted $\hat{\beta}$ is
\begin{eqnarray*}
\hat{\beta} =\frac{\sum_{i=1}^nY_i(x_i-\bar{x})}{\sum_{i=1}^n(x-\bar{x})^2}.
\end{eqnarray*}
When $Y_i$ has normal distribution,it is known $\hat{\beta}$ has normal distribution\cite{ref4}.

When $\rho=0$, we denote $X$ and $Y$ are independent. A test of independence of $X$ and $Y$ can be performed.

Let $r$ be the correlation coefficient, given $X_1=x_1, X_2=x_2, \dots, X_n=x_n(n>2)$, under the hypothesis $\rho=0$,
\begin{eqnarray*}
t=\frac{r\sqrt{n-2}}{\sqrt{1-r^2}}
\end{eqnarray*}
has a $t$-distribution\cite{ref3}\cite{ref4}. Using $t$-value we can evaluate the hypothesis $\rho=0$. In addition to test, we can perform another analysis by using mapping above defined.
When we apply statistical test of independence to real world data, we often consider the relation between two variables $\{x_i\}$ and $\{y_i\}$. We regard $\{x_i\}$ as explanatory variable and $\{y_i\}$ as objective variable. In analysis, $x_k$ a component of $\{x_i\}$ corresponds with $y_k$ a component of $\{y_i\}$, i.e. index $k$ is shared. We call pair of sequence $(\{x_i\},\{y_i\})$ original data. In original data, the correspondence represented $i$ defines the relation called exposure and outcome between two variables. For example, salt intake and blood pressure can be regarded  as exposure and outcome respectively. We considered that, besides analysis to original data, the analyses to $(\{x_{I^1_j}\},\{y_i\}),(\{x_{I^2_j}\},\{y_i\}),\dots,(\{x_{I^N_j}\},\{y_i\})$ might be meaningful. We call sets $(\{x_{I^1_j}\},\{y_i\}),\dots,(\{x_{I^N_j}\},\{y_i\})$ restricted "Random Data(RD)" or "Random Family (RF)". There are three advantages in constructing restricted random data. First, we can perform analysis in the context of non-exposure and compare exposure group with non-exposure group. Second, we can construct a number of restricted random data and get the distribution of a statistic (e.g. coefficient of least squares method $\hat{\beta}$ or correlation coefficient $\rho$ in the independence test) because random data is constructed via non-biased permutations: we can perform one-side or two-side test as in the permutation method. We call this test "Random Family Method(RFM)". Third, we can perform this method without losing independence of variables. Fig.\ref{fig:fig1} shows the scheme of this method. However, there are some disadvantages. First, we have to be careful in choosing exposure and outcome because interpretation of result can be complex. Second, restriction may cause bias because this restriction completely excludes the original order. But in terms of second disadvantage, the bias can be interpreted the effect we are interested in. We consider this method may be applied to inter-generational relation because we can confirm the inter-generational relation in the context of non-genetic relation.

\vspace{5mm}
\begin{figurehere}
  \begin{center}
    \includegraphics[width=11cm]{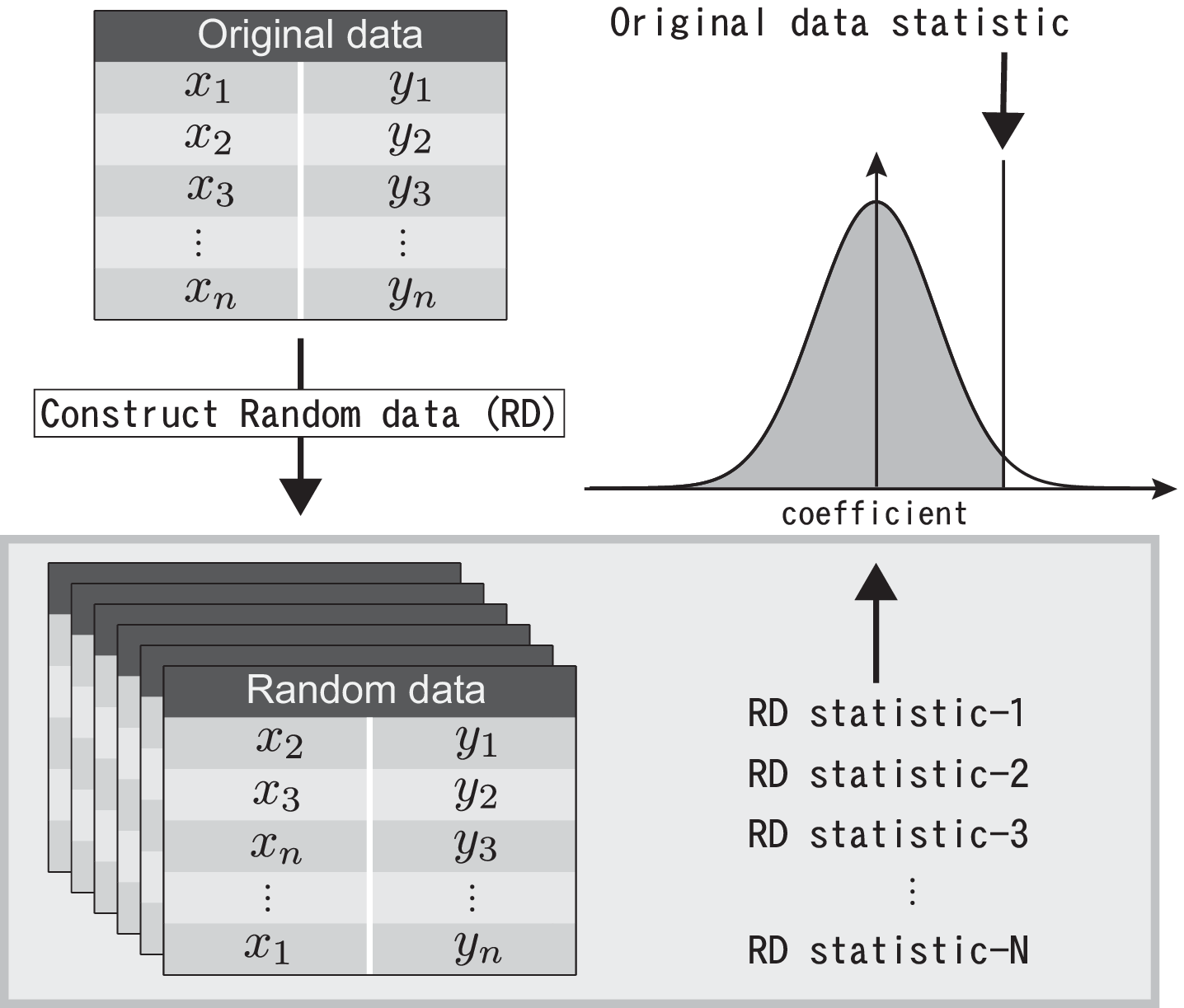}
\vspace{0mm}

 \caption{Scheme of Restricted re-sampling method. Construct restricted random data and calculate statistics. Using the distribution we can perform one-side or two-side test.}
    \label{fig:fig1}
  \end{center}
\end{figurehere}
\textbf{Case: }$\bm{N=10}$
\vspace{2mm}

We demonstrated the $N=10$ case in this section. Two sequence $\{x_i\}$ and $\{y_i\}$ was prepared. $\{x_i\}$ and $\{y_i\}$ has 10 components respectively.
\begin{eqnarray*}
\{x_i\}: (1,2,3,4,5,6,7,8,9,10) \\
\{y_i\}: (1,2,3,4,5,6,7,8,9,10)
\end{eqnarray*}
Fifth examples of random data for $\{x_i\} (N=10)$ are showed below.
\begin{eqnarray*}
\{x_{I^1_j}\}:(2,1,4,3,6,5,8,7,10,9)\\
\{x_{I^2_j}\}:(2,1,4,3,6,5,8,9,10,7)\\
\{x_{I^3_j}\}:(2,1,4,3,6,5,8,10,7,9)\\
\{x_{I^4_j}\}:(2,1,4,3,6,5,9,7,10,8)\\
\{x_{I^5_j}\}:(2,1,4,3,6,5,9,10,7,8)
\end{eqnarray*}
We calculated coefficients of least squares method $\hat{\beta}$ to $\{x_i\}$ for $N(10)=1,334,961$ times.\footnote{\noindent For $n>3$, $N(n)$ satisfies $N(n)=(n-1)(N(n-1)+N(n-2))$. $N(1)=0, N(2)=1, N(3)=2, N(4)=9, N(5)=44, N(6)=265, N(7)=1,854, N(8)=14,833, N(9)=133,496, N(10)=1,334,961$.} Fig.\ref{fig:fig2} shows the distribution of $\hat{\beta}$. The  mean of $\hat{\beta}$ was $-0.111$ and standard deviance was $0.315$. Difference of the mean from $0$ may be an effect of restricted re-sampling. An interpretation of such bias can be changed depending on properties of data used. This distribution follows normal distribution by Sapiro-Wilk test\cite{ref5}$(p<0.0001)$. In Fig.\ref{fig:fig2}，vertical line($\hat{\beta}=1$) shows the coefficient of original data. This line located out of 97.5\% cumulative area of the distribution. \footnote{In all analyses we used python 3.6.2. The details of python packages and algorithms used in analyses can be found in each website.}

\vspace{-0.5cm}
\begin{figurehere}
  \begin{center}
    \includegraphics[width=9cm]{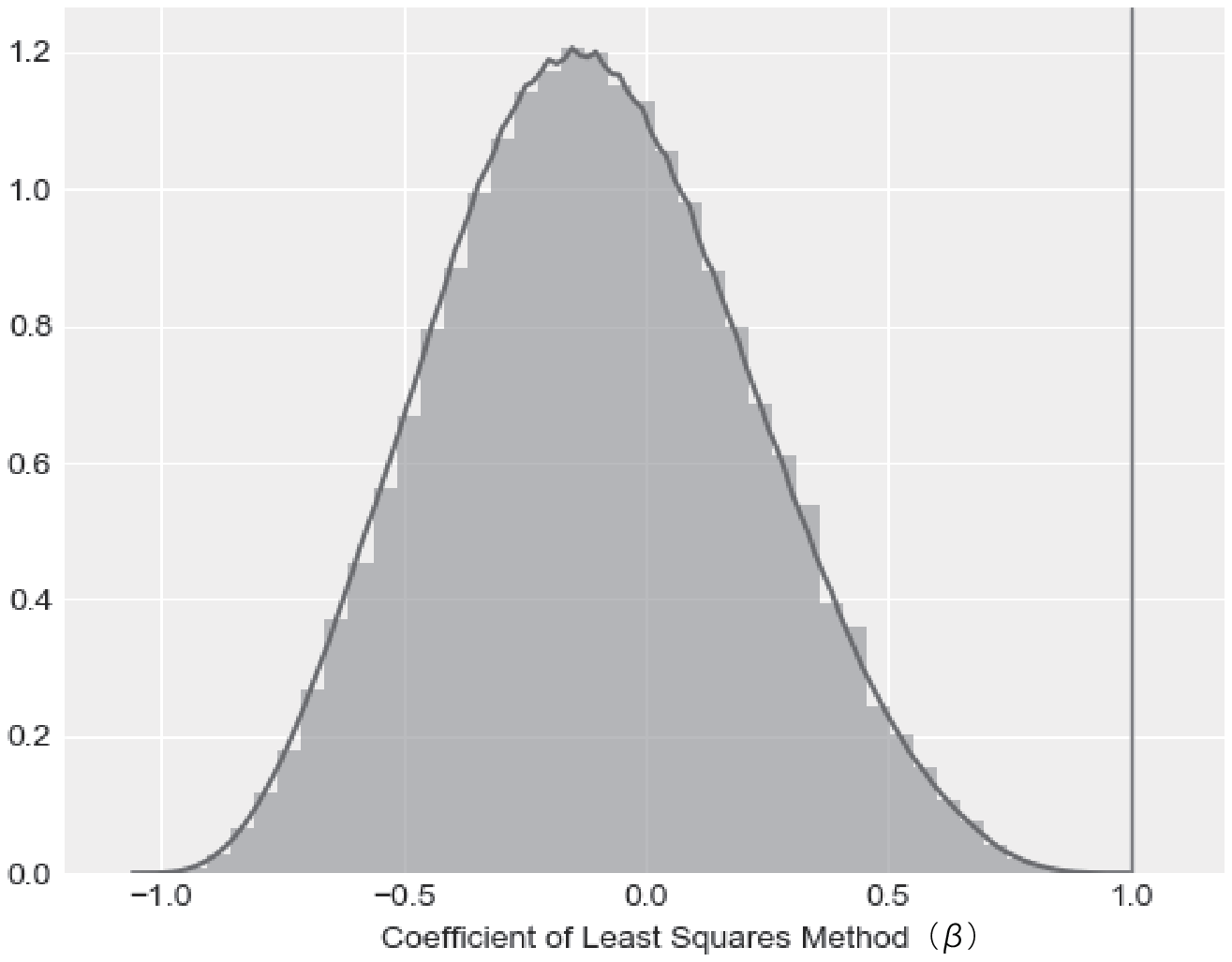}

    \caption{ Distribution of coefficients calculated by least squares method. Vertical axis is value of Kernel density estimation (KDE)\cite{ref6}. Curve in the figure is drawn by KDE. Vertical line in the figure shows $1$ which is the coefficient of original data $\{x_i\}, \{y_i\}$.}
 \label{fig:fig2}
  \end{center}
\end{figurehere}

\textbf{Conclusion}:We introduced a method to confirm inter-generational relations. This is an application of permutation method or bootstrap method: random family method.

\end{multicols}

\end{document}